\documentclass{article}

\usepackage{arxiv}

\usepackage[utf8]{inputenc} 
\usepackage[T1]{fontenc}    
\usepackage{hyperref}       
\usepackage{url}            
\usepackage{booktabs}       
\usepackage{amsfonts}       
\usepackage{amsmath}
\usepackage{nicefrac}       
\usepackage{microtype}      
\usepackage{lipsum}		
\usepackage{graphicx}
\usepackage{natbib}
\usepackage{doi}

\title{Hybrid SEIR–Temporal Fusion Transformer Framework with SHAP Analysis for Interpretable Pandemic Forecasting}

\author{ \href{https://orcid.org/0009-0004-5233-4499}{\includegraphics[scale=0.06]{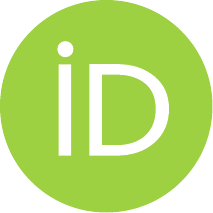}\hspace{1mm}Ashish Anton~Abraham} \\
	Department of Computer Science \& Engineering\\
	College of Engineering, Trivandrum, India\\
	\And
	\href{https://orcid.org/0009-0009-7078-3219}{\includegraphics[scale=0.06]{orcid.pdf}\hspace{1mm}Gopi Krishnan R} \\
	Department of Computer Science \& Engineering\\
	College of Engineering, Trivandrum, India\\
    \And
	\href{https://orcid.org/0009-0006-5232-7728}{\includegraphics[scale=0.06]{orcid.pdf}\hspace{1mm}Aiswarya Anil Nair} \\
	School of Sciences (SOS)\\
	Indira Gandhi National Open University, India\\
    \And
	\href{https://orcid.org/0009-0005-3026-0959}{\includegraphics[scale=0.06]{orcid.pdf}\hspace{1mm}Ajo Babu George}\thanks{Corresponding author: \href{mailto:drajo_george@dicemed.in}{drajo\_george@dicemed.in}. To know more about our research, visit: \url{https://www.dicemed.in/}.} \\ DiceMed \\ Cuttack, Odisha, India\\
    \And
	\href{https://orcid.org/0000-0002-5130-208X}{\includegraphics[scale=0.06]{orcid.pdf}\hspace{1mm}Kunal Agarwal} \\
	SCB Dental College \& Hospital\\
	Cuttack, Odisha, India\\
}

\renewcommand{\headeright}{Research Article}
\renewcommand{\undertitle}{Research Article}
\renewcommand{\shorttitle}{}

\hypersetup{
pdftitle={Hybrid SEIR--Temporal Fusion Transformer Framework with SHAP Analysis for Interpretable Pandemic Forecasting},
pdfsubject={cs.LG, cs.AI, q-bio.PE},
pdfauthor={Ashish Anton Abraham, Gopi Krishnan R, Aiswarya Anil Nair, Ajo Babu George, Kunal Agarwal},
pdfkeywords={Pandemic Forecasting, COVID-19, SEIR Model, Temporal Fusion Transformer, TFT, Explainable AI, SHAP, Epidemiological Modeling, Time Series Forecasting, Public Health Analytics},
}

\begin{document}
\maketitle

\begin{abstract}
	Forecasting pandemic spread across heterogeneous regions demands models that are both accurate and interpretable for public health action. This paper proposes a hybrid forecasting framework that integrates an SEIR epidemiological model enhanced with a Temporal Fusion Transformer (TFT) to produce interpretable multi-horizon infection forecasts. Using region-wise COVID-19 data from the Our World in Data corpus, we present a unified pipeline that integrates epidemiological indicators (susceptible, exposed, infectious, recovered), intervention signals (masking, vaccination, policies), mobility covariates, and demographic context. The SEIR component estimates latent disease dynamics and injects mechanistic priors, while the TFT learns non-linear temporal dependencies and cross-feature interactions from historical observations and known future inputs. To ensure transparency, SHAP (SHapley Additive exPlanations) is applied at both global and local levels, quantifying feature contributions and revealing region-specific drivers of surges and slowdowns. Compared with a calibrated standalone SEIR baseline, the hybrid SEIR-TFT approach provides more stable and accurate forecasts across epidemic phases, particularly under shifting intervention regimes. The framework is further operationalized through structured prediction and explanation outputs for interactive dashboards, enabling time-series, policy-impact, and geospatial analysis. By unifying mechanistic modeling, deep temporal learning, and explainable AI, this study offers a practical decision-support system for policy-sensitive pandemic forecasting.
\end{abstract}


\section{Introduction}
The COVID-19 pandemic demonstrated that infectious disease forecasting is not only a scientific challenge but also a policy-critical task. Health systems and governments require timely, region-specific predictions to allocate resources, plan interventions, and communicate risk effectively. Yet outbreak dynamics are shaped by shifting transmission behavior, vaccination progress, demographic heterogeneity, and policy actions, making reliable multi-region forecasting difficult. Models used in practice must therefore satisfy two requirements simultaneously: strong predictive performance under changing conditions and clear interpretability for high-stakes decision-making.

Classical compartmental epidemiological models, particularly the Susceptible-Exposed-Infected-Recovered (SEIR) framework, provide a standard representation of disease progression and remain valuable for mechanistic understanding. However, conventional SEIR formulations often rely on simplifying assumptions and fixed or weakly adaptive parameters, which can limit their ability to represent nonlinear temporal effects and complex cross-regional variability. In contrast, deep learning time-series models can learn rich temporal patterns from heterogeneous data, but they are frequently criticized for limited transparency. This creates a practical gap between mechanistic interpretability and data-driven forecasting power.

To bridge this gap, this study proposes a hybrid framework for interpretable pandemic forecasting. The approach combines an enhanced SEIR component, which estimates latent epidemiological states and supplies mechanistic priors, with a temporal fusion transformer (TFT), which performs multi-horizon forecasting by modeling interactions among static covariates, historical observations, and known future inputs. Region-wise COVID-19 data from Our World in Data are integrated with demographic factors and policy-related indicators to support robust multi-regional prediction.

A central objective of this work is explainability. To make forecasts actionable rather than opaque, SHapley Additive exPlanations (SHAP) are incorporated to quantify feature influence at both global and local levels. This enables identification of the most influential drivers behind predicted infection trends across regions and time windows, improving trust and supporting evidence-based intervention planning.

Beyond model development, the project includes an analytics pipeline that exports prediction and explanation outputs for interactive visualization. Time-series trend comparison, policy-impact exploration, and geospatial views are designed to translate model outputs into operational insights for stakeholders. In this way, the framework is positioned as both a forecasting engine and a decision-support tool.

The main contributions of this study are threefold:  
1. A hybrid epidemiological-learning architecture that integrates SEIR structure with TFT-based temporal modeling for multi-region pandemic forecasting.  
2. An explainable AI layer using SHAP to provide transparent, feature-level interpretation of model behavior.  
3. An end-to-end simulation and visualization workflow that links predictive modeling with policy-relevant, interactive analysis.
Overall, this work advances pandemic forecasting by unifying mechanistic epidemiology, transformer-based sequence learning, and explainable AI in a single interpretable framework aimed at practical public health use.

\section{Related Work}
\label{sec:relatedwork}

Epidemic forecasting literature can be broadly grouped into three streams: mechanistic compartmental modeling, data-driven time-series learning, and explainable AI for model transparency.

Classical compartmental models such as SI (Susceptible-Infected), SIR (Susceptible-Infected-Recovered) \citet{kermack1927contribution}, and SEIR (Susceptible-Exposed-Infected-Recovered) \citet{cooke1967functional} remain foundational in infectious disease analysis because they encode transmission processes in an interpretable mathematical framework. In particular, the SEIR model is widely used for diseases with an incubation or latent period, as it explicitly models individuals who have been exposed to the pathogen but are not yet infectious. These models are useful for scenario simulation and intervention reasoning. Still, their practical forecasting performance is often constrained by simplifying assumptions, including homogeneous mixing, fixed parameters, and limited responsiveness to evolving policy and behavioral changes across regions (\citet{10.1371/journal.pone.0298932}). As a result, purely mechanistic models struggle to capture heterogeneous multi-region dynamics under rapidly changing real-world conditions.

To improve predictive accuracy, recent studies have explored machine learning and deep learning methods for epidemic time-series forecasting. \citet{Zhu2019} has explored recurrent neural network (RNN) architectures to forecast influenza epidemic spread. \citet{8581423} demonstrates that LSTMs (Long Short-Term Memory) and GRUs (Gated Recurrent Unit) can effectively capture nonlinear temporal dynamics and spatio-temporal correlations in epidemiological data. Self-attention based transformer models (\citet{wu2020deeptransformermodelstime}) have been studied to be particularly effective in time series forecasting problems due to the attention mechanisms that better capture long-
range dependencies and feature interaction. Within transformer-based approaches, extensive studies like \citet{Kalahasti2025.02.24.25322795} evaluates diverse architectures \textit{TabPFN-TS}, \textit{TimeGPT}, \textit{TimesFM}, \textit{Lag-Llama}, and \textit{Chronos} across diverse pathogens, diseases, and locations, including influenza-like illness, RSV, chickenpox, dengue, COVID-19, and neonatal bronchiolitis. Despite their strong predictive performance, these neural forecasting frameworks generally provide limited interpretability, making it difficult to understand which epidemiological, temporal, or intervention-related factors drive model predictions and reducing transparency for public health decision-making.

Earlier approaches, such as \citet{guo2019exploringinterpretablelstmneural} tried to solve this \textit{'black box'} problem by enabling the LSTM to learn variable-wise hidden states, intending to capture different dynamics in multi-variable time series. Building on this direction, \citet{lim2019temporal} introduced the Temporal Fusion Transformer (TFT), which combined recurrent sequence modeling with attention-based feature selection, gating layers, and interpretable temporal attention to improve both predictive performance and transparency in multi-horizon forecasting tasks. Further research, including \citet{10224685}, has shown the effectiveness of using TFT for not only forecasting but also the interpretability part. Hybrid epidemiological–deep learning frameworks have also been explored that combine mechanistic compartmental modeling with neural sequence architectures. These include SEIR-LSTM (\citet{gan2021trend}) and ARIMA-LSTM hybrids (\citet{wang2024arima}), as well as transformer-enhanced mechanistic models such as \citet{11154197}. 

In SEIR-LSTM, the SEIR component estimates disease transmission dynamics, while the LSTM integrates auxiliary variables such as migration statistics and human mobility patterns to better model changing intervention conditions and regional spread behavior. The approach remains sensitive to bias in the SEIR estimates (\citet{gan2021trend}). Similarly, in the ARIMA-LSTM hybrid, ARIMA models linear temporal trends, and LSTM captures nonlinear epidemic dynamics. Although this reduces reliance on predefined epidemiological assumptions, the framework still suffers from limited epidemiological interpretability. Moreover, ARIMA components perform poorly under highly nonlinear transmission dynamics (\citet{wang2024arima}), and LSTMs have an inherent limitation of failing to capture long-range dependencies.

In contrast, the SEIR-Informer hybrid demonstrates stronger long-sequence modeling and global dependency learning capabilities, achieving lower prediction errors than ARIMA-LSTM while remaining more responsive to epidemic inflection points caused by policy interventions and behavioral changes. Nevertheless, the framework works on city-level data and also exhibited overestimation in later outbreak stages when latent social interventions and behavioral adaptations are difficult to quantify, potentially leading to overfitting under highly volatile sequences (\citet{11154197}).

Parallel to forecasting advances, explainability has become a critical requirement in public health AI. High-performing black-box models are often insufficient for policy settings where stakeholders need to understand why forecasts change across time and place. SHapley Additive exPlanations (SHAP) (\citet{lundberg2017unifiedapproachinterpretingmodel}) has emerged as a widely used post-hoc explainability method because it provides consistent feature-attribution analysis at both global and local levels. In epidemic applications, SHAP helps identify which factors (e.g., vaccination trends, intervention indicators, demographic context) contribute most to predicted surges or declines, improving trust and supporting actionable interpretation. Beyond tabular feature attribution methods such as SHAP, explainable AI has also been applied in clinical imaging workflows, where visual attribution techniques such as Grad CAM and Grad CAM++ have been used to highlight diagnostically relevant regions in deep learning based oral cancer detection systems \citep{george2025grad}. These studies reinforce the broader importance of transparent model outputs for clinical trust and adoption, although the present work focuses on temporal and feature level explanations for epidemiological forecasting rather than image based diagnosis.

Many studies optimize either mechanistic credibility or predictive flexibility, while fewer integrate both together with systematic explainability and deployment-oriented analytics. The proposed architecture addresses this gap through a hybrid SEIR-TFT framework with SHAP-based annotations. SEIR-derived latent epidemiological signals provide domain-grounded structure, TFT captures nonlinear multi-region temporal dependencies, and SHAP offers transparent feature-level explanations. This combination positions the framework as a practical bridge between epidemiological interpretability, forecasting accuracy, and decision-support usability.

\section{Methodology}
\subsection{SEIR}
The SEIR model is a widely used method to simulate transmission dynamics of infectious diseases like COVID-19, SARS, and influenza which involve incubation periods. SEIR includes this transition period in the calculation from \textit{susceptible} to \textit{infected}. Once individuals come in contact with the pathogen, they move from \textit{susceptible} to \textit{exposed} state. After the incubation period, the individual transitions from \textit{exposed} to \textit{infected} state and later to \textit{recovered} state after treatment and care.
\begin{equation}
	\frac{dS}{dT}=-\frac{hIS}{n}
\end{equation}
\begin{equation}
	\frac{dE}{dt}=\frac{hIS}{N}-iE
\end{equation}
\begin{equation}
	\frac{dI}{dt}=iE-rI-kI
\end{equation}
\begin{equation}
	\frac{dR}{dt}=rI
\end{equation}
\begin{equation}
	\frac{dD}{dt}=kI
\end{equation}
Applying in real-world scenarios, each latent state is calculated from statistics and parameters reported by the country's official body. For COVID-19, the incubation period is generally considered to be around 5 days. Epidemiological estimates suggest that a single infected individual can transmit the disease to an average of 13.8 susceptible individuals while exposed, but pre-symptomatic individuals may also infect nearly three susceptible persons (\citet{zheng2020prediction}). The daily cumulative number of infected cases is computed by combining newly confirmed infections with recovered cases. Furthermore, as healthcare facilities catch up during the later phase of the outbreak, the recovery rate parameter was adjusted to account for the increased medical capacity and improved patient management.
\subsection{Temporal Fusion Transformer}

The \textbf{Temporal Fusion Transformer (TFT)} is a deep learning architecture introduced in \citet{lim2019temporal} for interpretable multi-horizon time-series forecasting. By integrating recurrent sequence modeling with transformer-based attention mechanisms, TFT is capable of capturing both \textit{short-term temporal variations} and \textit{long-range dependencies} in sequential data. Its architecture is particularly suitable for epidemic forecasting because it can simultaneously process heterogeneous inputs while also providing interpretability through attention weights.

\subsubsection{Architecture}

The model processes three categories of input variables:

\begin{itemize}
	\item \textbf{Static covariates}: Each spatial entity \(i\) is associated with static features \(s_i\), representing time-invariant demographic and geographic attributes.

	\item \textbf{Observed historical inputs}: Time-dependent observed variables are denoted by \(z_{i,t}\), while target observations are represented as \(y_{i,t}\). For a lookback window of length \(k\), the historical sequences are defined as \(y_{i,t-k:t}=\{y_{i,t-k},\dots,y_{i,t}\}\) and \(z_{i,t-k:t}=\{z_{i,t-k},\dots,z_{i,t}\}\).

	\item \textbf{Known future inputs}: Future-known covariates are represented as \(x_{i,t}\), where values remain available throughout the forecasting horizon \(\tau\). The corresponding sequence is given by \(x_{i,t-k:t+\tau}=\{x_{i,t-k},\dots,x_{i,t+\tau}\}\).
\end{itemize}

Initially, all inputs are transformed into latent vector representations through embedding and input transformation layers. These transformed features are then passed through a Variable Selection Network (VSN), which identifies the most informative variables while reducing the influence of noisy or redundant features. This mechanism improves both forecasting accuracy and model interpretability.

\subsubsection{Temporal Modeling Using LSTM}

After feature selection, TFT employs LSTM networks to model local temporal dependencies. The encoder receives historical observations, whereas the decoder processes future-known inputs. This encoder--decoder configuration enables the model to effectively capture local temporal structures such as trends, seasonal patterns, and abrupt fluctuations.

Static contextual information is also incorporated into the temporal pathway to guide feature learning. Such contextual enrichment allows the model to adapt temporal representations according to region-specific or population-specific characteristics, which is particularly important in epidemic forecasting scenarios.

\begin{figure}[t]
    \centering
    \includegraphics[width=0.58\textwidth]{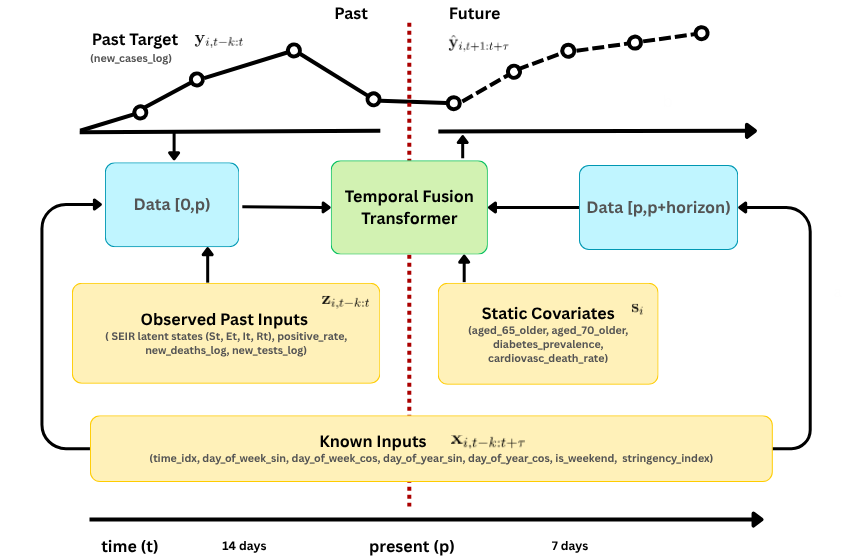}
    \caption{Time-series forecasting using Temporal Fusion Transformer}
    \label{fig:architecture}
\end{figure}

\subsubsection{Static Enrichment Layer}

To further improve contextual understanding, TFT introduces a static enrichment layer based on \textit{Gated Residual Networks (GRNs)}. This layer combines static metadata with temporal feature representations, enabling the model to learn how fixed contextual factors influence evolving temporal dynamics.

The integration of static and temporal information enhances the model's ability to represent complex disease transmission behavior influenced by demographic conditions, healthcare infrastructure, or social characteristics.

\subsubsection{Interpretable Self-Attention Mechanism}

One of the primary strengths of TFT lies in its interpretable \textbf{ multi-head self-attention} mechanism. Unlike traditional recurrent models that mainly focus on nearby observations, the attention mechanism can selectively assign importance to relevant historical time steps across the entire sequence.

The standard attention operation is defined as:

\begin{equation}
	\mathrm{Attention}(Q,K,V)=A(Q,K)V
\end{equation}

where \(Q\), \(K\), and \(V\) denote the \textit{query}, \textit{key}, and \textit{value} matrices, respectively, and \(A(\cdot)\) represents the normalization function used to compute attention scores.

To improve representational capacity, TFT adopts a \textbf{multi-head attention} mechanism, where multiple attention heads learn different temporal relationships simultaneously:

\begin{equation}
	\mathrm{MultiHead}(Q,K,V)=[H_1,\dots,H_m]W_H
\end{equation}

Here, \(W_H\) denotes the learnable projection matrix used to combine the outputs of multiple attention heads $\{H_1,\ldots,H_m\}$ into a unified representation. Each attention head is computed as:

\begin{equation}
	H_h=\mathrm{Attention}(QW_Q^{(h)},KW_K^{(h)},VW_V^{(h)})
\end{equation}

\(W_Q^{(h)}\), \(W_K^{(h)}\), and \(W_V^{(h)}\) represent the learnable projection matrices for the query, key, and value representations associated with the \(h\)-th attention head. To enhance interpretability, TFT modifies the conventional transformer attention mechanism by sharing value representations across all heads and aggregating them additively:

\begin{equation}
	\mathrm{InterpretableMultiHead}(Q,K,V)=\tilde{H}W_H
\end{equation}

where

\begin{equation}
	\tilde{H}=\frac{1}{H}\sum_{h=1}^{m_H}
	\mathrm{Attention}(QW_Q^{(h)},KW_K^{(h)},VW_V)
\end{equation}

This formulation improves transparency by enabling direct interpretation of temporal importance across forecasting horizons.

\subsection{Attention Weight Representation}

Following the static enrichment stage, temporal feature vectors are organized into a sequence matrix containing both encoder inputs and forecasting horizons. The self-attention layer computes attention weights that quantify the contribution of previous observations to future predictions.

The temporal attention representation for forecasting horizon \(\tau\) is expressed as:

\begin{equation}
	\beta(t,\tau)=\sum_{n=-k}^{\tau_{\max}}
	\alpha(t,n,\tau)\tilde{\theta}(t,n)
\end{equation}

where \(t\) denotes the current reference time at which the forecast is generated, \(\alpha(t,n,\tau)\) represents the attention score assigned to time step \(n\) when predicting horizon \(\tau\), and \(\tilde{\theta}(t,n)\) is the transformed temporal feature vector. Larger attention values indicate stronger influence of historical observations on future forecasts.

\section{Experimental Setup}
\subsection{Dataset and Preprocessing}

This study utilizes the \textbf{Our World in Data (OWID) COVID-19} dataset, a large-scale public repository containing daily country-level COVID-19 statistics compiled from multiple official and institutional sources, including Johns Hopkins University (JHU), the Oxford COVID-19 Government Response Tracker (OxCGRT), the World Bank, and the United Nations. The dataset follows a panel time-series structure with one record per country and date, covering epidemiological trends, testing activity, policy interventions, demographic indicators, and healthcare-related statistics \citet{owid-coronavirus}.

To construct the forecasting framework, we retain three categories of features:

\begin{itemize}
	\item \textbf{Core temporal variables}: epidemiological and intervention-related indicators, including \texttt{new\_cases}, \texttt{new\_deaths}, \texttt{new\_tests}, \texttt{positive\_rate}, and \texttt{stringency\_index}.
	
	\item \textbf{Static demographic variables}: country-level contextual covariates such as \texttt{aged\_65\_older}, \texttt{aged\_70\_older}, \texttt{diabetes\_prevalence}, \texttt{cardiovasc\_death\_rate}, and \texttt{population}.
	
	\item \textbf{Temporal index variables}: calendar-derived cyclic encodings and a weekend indicator used to capture periodic reporting patterns.
\end{itemize}

To reduce structural bias, aggregated entities such as \textit{World}, continental groups, and income-based categories were excluded from the dataset. The data was filtered beginning from \textbf{2020-03-01} and sorted chronologically for each country. Missing temporal covariates were harmonized using forward and backward filling strategies, while missing \texttt{new\_cases} observations were assigned a value of zero.

\subsubsection{Temporal Smoothing and Transformation}

COVID-19 reporting often contains strong weekly fluctuations caused by testing delays and reporting inconsistencies. To reduce short-term noise, a 7-day moving average was applied to the daily case counts:

\begin{equation}
	\mathrm{new\_cases\_7d\_avg}_t=
	\frac{1}{7}\sum_{i=0}^{6}\mathrm{new\_cases}_{t-i}
\end{equation}

Let \(C_t\) denote the smoothed epidemic signal:

\begin{equation}
	C_t=\mathrm{new\_cases\_7d\_avg}_t
\end{equation}

A logarithmic transformation was further applied to stabilize variance and reduce the influence of extreme outbreak spikes:

\begin{equation}
	x_t^{\log}=\log(1+x_t)
\end{equation}

\subsubsection{SEIR Latent-State Approximation}

Rather than explicitly solving differential equations, latent SEIR compartment states were approximated directly from observed epidemiological data. This approach enables efficient integration of mechanistic epidemic priors into the forecasting pipeline.

The infected population ratio was estimated using a rolling 14-day infectious window:

\begin{equation}
	I_t=
	\frac{\sum_{i=1}^{13}C_{t-i}}
	{\mathrm{population}}
\end{equation}

The recovered population proxy was computed from cumulative infections excluding the active infectious population:

\begin{equation}
	R_t=
	\frac{
	\left(\sum_{r=1}^{t}C_r\right)
	-
	\left(\sum_{i=0}^{13}C_{t-i}\right)
	}
	{\mathrm{population}}
\end{equation}

The exposed compartment was approximated using recent infections scaled by an incubation-related factor:

\begin{equation}
	E_t=
	\frac{
	2.5\sum_{i=0}^{4}C_{t-i}
	}
	{\mathrm{population}}
\end{equation}

The susceptible population was then derived as:

\begin{equation}
	S_t=
	1-(E_t+I_t+R_t)
\end{equation}

To ensure numerical consistency, all compartment values were clipped to the interval \([0,1]\) and normalized such that:

\begin{equation}
	S_t+E_t+I_t+R_t=1
\end{equation}

Static covariates were median-imputed independently for each country and subsequently standardized. Temporal variables were normalized using statistics computed exclusively from the training period to avoid information leakage. The scaler fitting cutoff was fixed at 2021-12-31 and later applied consistently across validation and testing intervals.

For sequential forecasting, a country-specific temporal index was constructed. The experiments were conducted using four temporally separated evaluation splits corresponding to different pandemic phases, including the primary period, rising third-wave phase, falling third-wave phase, and post-third-wave period, as shown in Table~\ref{tab:data_splits}.

\begin{table}[h]
	\caption{Temporal data splits used for model training \& evaluation.}
	\centering
	\begin{tabular}{llllll}
		\toprule
		\textbf{Split} & \textbf{Train End} & \textbf{Val Start} & \textbf{Val End} & \textbf{Test Start} & \textbf{Test End} \\
		\midrule
		Primary & 29-11-2021 & 30-11-2021 & 14-12-2021 & 15-12-2021 & 29-12-2021 \\
		
		Rising 3rd Wave & 31-12-2021 & 01-01-2022 & 15-01-2022 & 16-01-2022 & 30-01-2022 \\
		
		Falling 3rd Wave & 31-01-2022 & 01-02-2022 & 15-02-2022 & 16-02-2022 & 02-03-2022 \\
		
		Post 3rd Wave & 28-02-2022 & 01-03-2022 & 15-03-2022 & 16-03-2022 & 30-03-2022 \\
		\bottomrule
	\end{tabular}
	\label{tab:data_splits}
\end{table}

\subsection{Hybrid SEIR-TFT Integration}
The forecasting architecture combines mechanistic SEIR-derived epidemic representations with the Temporal Fusion Transformer (TFT) to jointly model epidemiological structure and nonlinear temporal dynamics. The latent SEIR compartment states \((S_t, E_t, I_t, R_t)\), computed from rolling normalized infection statistics, are incorporated as \texttt{time\_varying\_unknown\_reals} within the TFT framework. 
The final model inputs are organized into four categories:

\begin{itemize}
	\item \textbf{Static real-valued covariates}: country-specific demographic variables that remain constant over time, including \texttt{aged\_65\_older}, \texttt{aged\_70\_older}, \texttt{diabetes\_prevalence}, and \texttt{cardiovasc\_death\_rate}.
	
	\item \textbf{Time-varying known real-valued inputs}: temporally indexed features available throughout the forecasting horizon, including \texttt{time\_idx}, \texttt{day\_of\_week\_sin}, \texttt{day\_of\_week\_cos}, \texttt{day\_of\_year\_sin}, \texttt{day\_of\_year\_cos}, \texttt{is\_weekend}, and \texttt{stringency\_index}.
	
	\item \textbf{Time-varying unknown real-valued inputs}: historical observations unavailable in the future, consisting of the target variable \texttt{new\_cases\_log}, SEIR latent states \((S_t,E_t,I_t,R_t)\), \texttt{positive\_rate}, \texttt{new\_deaths\_log}, and \texttt{new\_tests\_log}.
	
	\item \textbf{Sequence identifier}: the grouped temporal index \texttt{location\_idx}, which separates country-specific trajectories during multi-horizon forecasting.
\end{itemize}
This integration injects epidemiological priors directly into the sequence model while preserving the flexibility of attention-based temporal learning. For model training, each input sequence consists of an encoder window containing up to 14 historical time steps (with a minimum context length of 7), and the model predicts the subsequent 7-day forecasting horizon. An illustration is shown in Figure~\ref{fig:architecture}.

\subsection{Model Training}
The TFT was trained using a probabilistic forecasting objective based on quantile regression. Instead of predicting a single deterministic value, the model estimates multiple conditional quantiles of the target distribution, enabling uncertainty-aware forecasting. Specifically, the model optimized \texttt{QuantileLoss([0.1, 0.5, 0.9])}, enabling simultaneous prediction of the 10th, 50th, and 90th conditional quantiles corresponding to lower-bound, median, and upper-bound forecasts. Together, the \(0.1\) and \(0.9\) quantiles define an approximate \(80\%\) prediction interval for future COVID-19 case estimates.
The quantile loss function for a target value \(y_t\), predicted quantile \(\hat{y}_t^{(q)}\), and quantile level \(q \in (0,1)\) is defined as:
\begin{equation}
	L_q(y_t,\hat{y}_t^{(q)})=\max \left(q(y_t-\hat{y}_t^{(q)}),(q-1)(y_t-\hat{y}_t^{(q)})\right)
\end{equation}

\begin{figure}[t]
    \centering
    \includegraphics[width=0.58\textwidth]{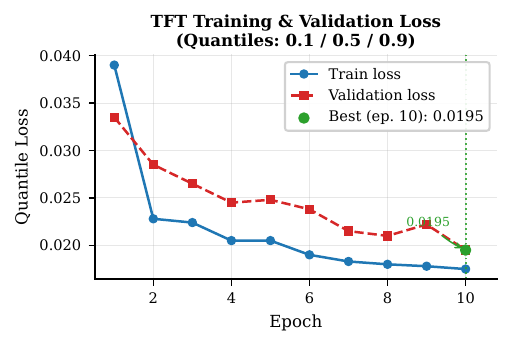}
    \caption{Training and validation quantile loss curves for the Temporal Fusion
    Transformer (TFT) model trained at quantiles $\{0.1, 0.5, 0.9\}$.
    The model converges steadily over 10 epochs, with the best validation loss
    of $0.0195$ achieved at epoch~10 (dotted vertical line).
    The gap between training and validation loss narrows across epochs,
    indicating stable generalisation without overfitting.}
    \label{fig:tft_loss}
\end{figure}

Hyperparameter optimization was conducted using Optuna with Bayesian search over 50 trials. The search space included learning rate \((10^{-4} \text{ to } 10^{-2})\), hidden size \((16\text{--}128)\), hidden continuous size \((8\text{--}64)\), attention head size \((1\text{--}4)\), dropout rate \((0.1\text{--}0.5)\), and gradient clipping value \((0.01\text{--}1.0)\). The final optimized configuration achieved a best validation loss of 0.0195 (Figure~\ref{fig:tft_loss}) with learning rate \(=0.00618\), hidden size \(=52\), hidden continuous size \(=17\), attention head size \(=1\), dropout \(=0.446\), and gradient clipping value \(=0.222\).

Training employed early stopping with a patience of five epochs to reduce overfitting. All experiments were performed independently across the four phase-aware temporal evaluation splits.

\section{Results and Analysis}
\subsection{Explainability Analysis}
For healthcare applications, explainability is critical since accurate predictions without transparent reasoning mechanisms do not help policy deployment. Epidemic forecasts must therefore be interpretable and should enable evidence-driven actions. To obtain feature-level attribution and transparency, we have employed \textbf{SHapley Additive exPlanations (SHAP)}, a game-theoretic explainability framework that estimates the marginal contribution of each feature toward the final forecast \citet{lundberg2017unifiedapproachinterpretingmodel}. SHAP was selected because it provides both global and local interpretability, enabling analysis of overall feature importance as well as instance-specific prediction behavior across different countries and outbreak phases.
\subsubsection{SHAP-Based Feature Attribution}
\begin{figure}[t]
    \centering
    \includegraphics[width=\textwidth]{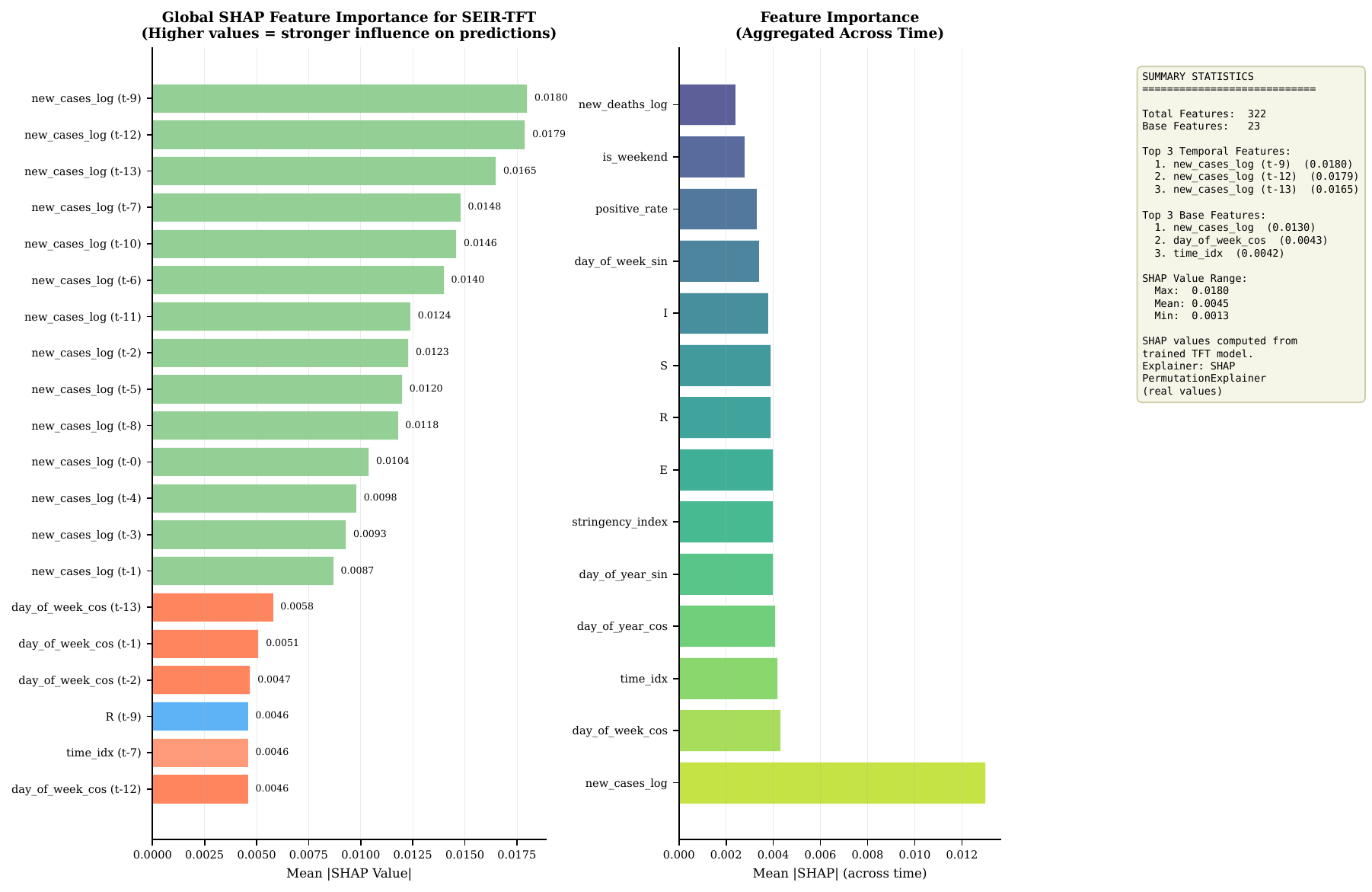}
    \caption{
    Global SHAP feature importance analysis for the proposed SEIR-TFT framework. 
    The top panel shows timestep-specific feature attributions, highlighting the dominant influence of historical infection trajectories such as \texttt{new\_cases\_log (t-9)} and \texttt{new\_cases\_log (t-13)}. 
    The bottom-left panel presents feature importance aggregated across temporal windows, where \texttt{new\_cases\_log}, \texttt{day\_of\_week\_cos}, and \texttt{time\_idx} emerge as the most influential base features. 
    The bottom-right panel summarizes overall SHAP statistics computed using the trained TFT model.
    }
    \label{fig:seir_tft_global_importance}
\end{figure}

SHAP explanations identified the dominant epidemiological and intervention-related factors influencing forecast trajectories. A model-agnostic SHAP explainer was applied to the trained TFT model using 200 sampled sequences from the validation set. The resulting SHAP matrix contained \(322\) flattened temporal features spanning all encoder timesteps and covariates. Global feature importance was computed using the mean absolute SHAP value across all samples. Figure~\ref{fig:seir_tft_global_importance} presents the overall feature ranking and aggregated feature contributions across time.

The SHAP analysis revealed that historical infection trajectories were the dominant drivers of forecasting behavior. In particular, lagged epidemiological signals such as \texttt{new\_cases\_log (t-9)}, \texttt{new\_cases\_log (t-12)}, and \texttt{new\_cases\_log (t-13)} achieved the highest attribution scores, indicating that the model relies strongly on medium-range historical case dynamics when predicting future outbreaks. This behavior is consistent with the autoregressive nature of epidemic spread, where recent transmission patterns provide critical information about near-future infection trends.

At the aggregated feature level, \texttt{new\_cases\_log} emerged as the most influential base feature overall, followed by temporal indicators such as \texttt{day\_of\_week\_cos}. The prominence of cyclical temporal features suggests that the TFT successfully captures reporting periodicities and weekly behavioral patterns commonly observed in epidemiological data.
Notably, \texttt{stringency\_index} ranked among the most influential non-epidemiological variables, indicating high relevance of government intervention measures to forecasting outcomes. Overall, these findings demonstrate the TFT model's ability to capture short-term temporal dependencies, medium-range epidemic memory, long-term temporal evolution, and the impact of public health interventions simultaneously.

\begin{figure}[h]
    \centering
    \includegraphics[width=0.8\textwidth]{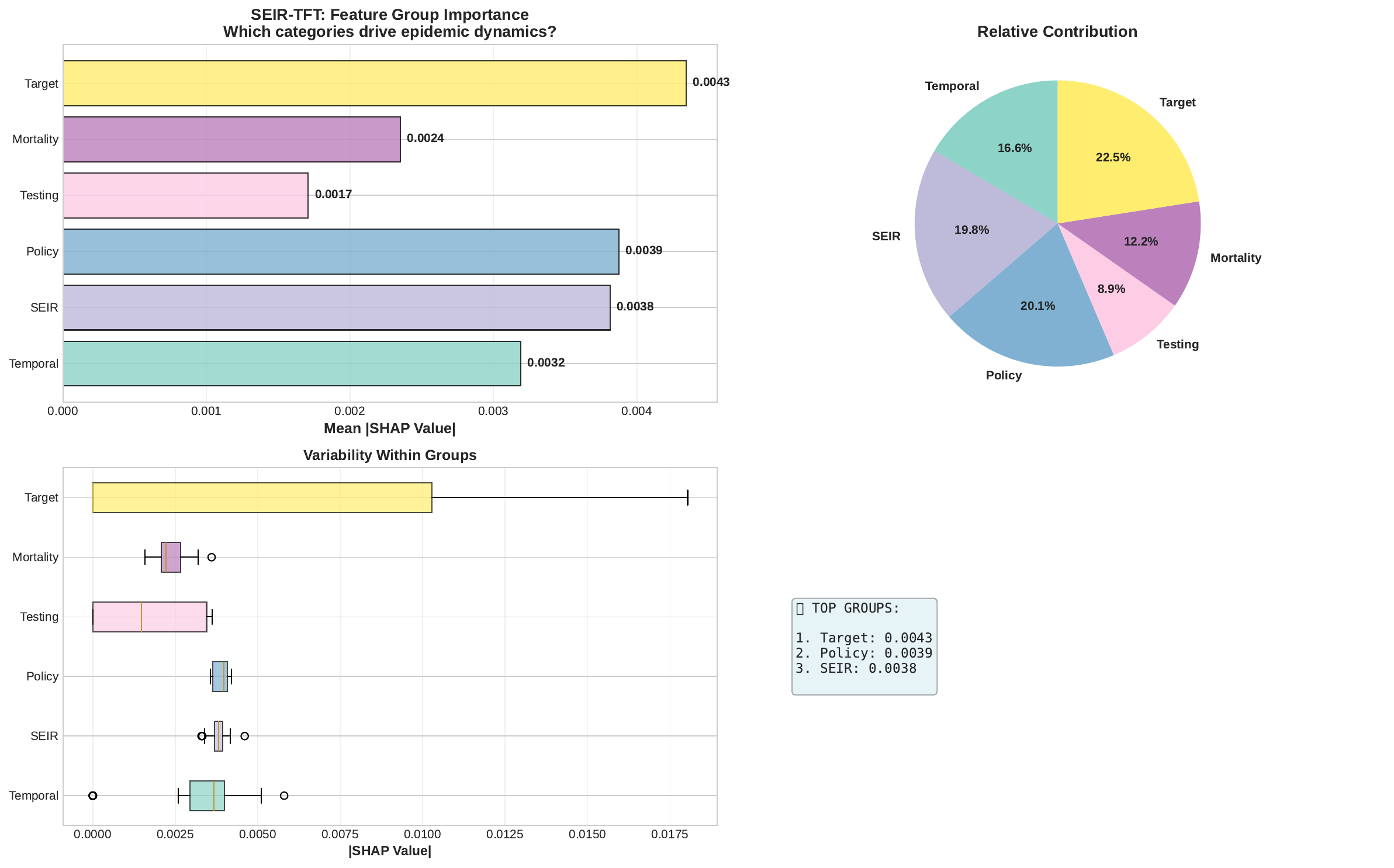}
    \caption{SEIR vs TFT+SEIR overall performance comparison. Target features, policy interventions \& SEIR latent states lead transmission dynamics.}
    \label{fig:seir-tft_feature_groups}
\end{figure}

Feature-group attribution further showed that multiple information sources contributed meaningfully to forecasting performance. As illustrated in Figure~\ref{fig:seir-tft_feature_groups}, target-history variables accounted for approximately \(22.5\%\) of the total contribution, followed by policy-related variables (\(20.1\%\)), SEIR latent states (\(19.8\%\)), temporal encodings (\(16.6\%\)), mortality indicators (\(12.2\%\)), and testing-related variables (\(8.9\%\)). The relatively high contribution of policy and SEIR groups suggests that the hybrid architecture successfully integrates mechanistic epidemic dynamics with external intervention signals.

\begin{figure*}[t]
    \centering
    \includegraphics[width=\textwidth]{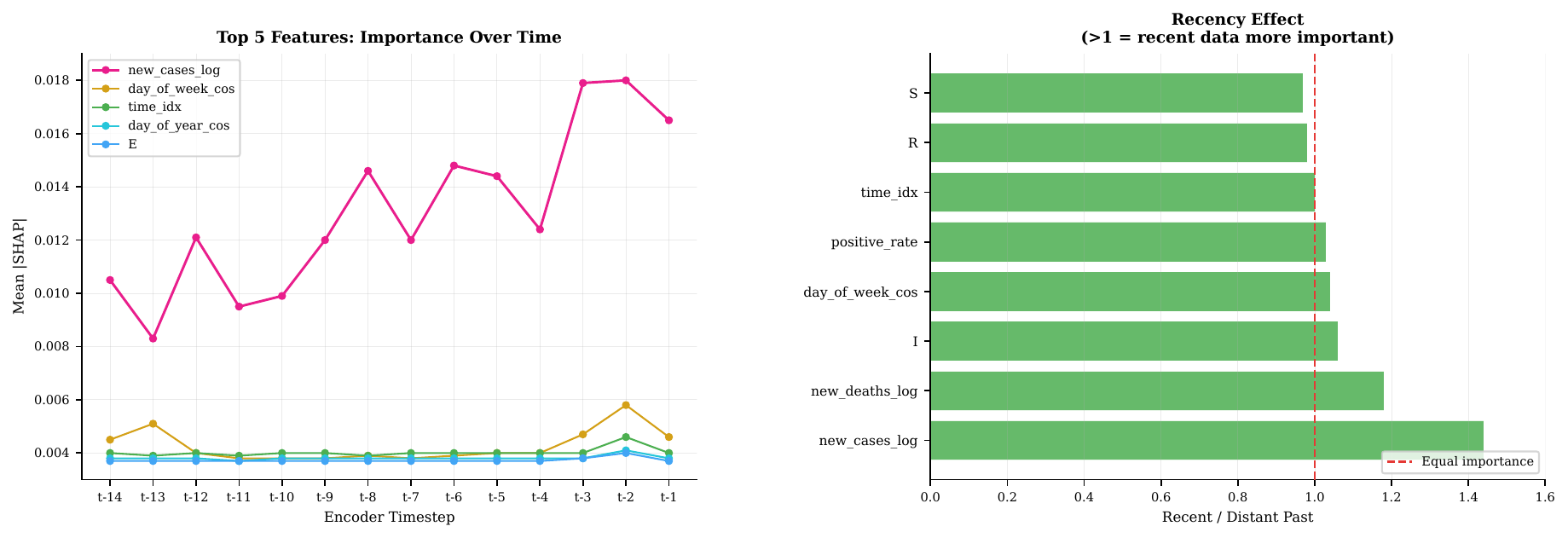}
    \caption{
    Temporal SHAP importance analysis for the proposed SEIR-TFT forecasting framework. 
    The top-left panel illustrates the evolution of feature importance across encoder timesteps, demonstrating that historical epidemiological signals maintain strong influence throughout the temporal window. 
    The top-right panel shows aggregate timestep importance, revealing progressively higher attribution scores for observations closer to the prediction horizon, consistent with autoregressive epidemic dynamics. 
    The bottom-left panel tracks temporal importance trajectories for the most influential features, where \texttt{new\_cases\_log} remains dominant across nearly all encoder positions. 
    The bottom-right panel presents the recency-effect analysis, comparing recent and distant historical contributions, indicating that the model prioritizes recent epidemiological context while still preserving long-range temporal dependencies.
    }
    \label{fig:seir_tft_temporal_importance}
\end{figure*}

Temporal SHAP analysis further revealed that encoder timesteps closer to the prediction horizon exhibited progressively higher attribution scores, indicating that recent epidemic observations were the most informative for short-term forecasting. As illustrated in Figure~\ref{fig:seir_tft_temporal_importance}, the model assigns increasing importance from distant historical context (\(t-14\)) toward more recent observations (\(t-1\)), consistent with the autoregressive nature of epidemic transmission dynamics. 

The temporal attribution trajectories additionally demonstrate that historical infection signals, particularly \texttt{new\_cases\_log}, remain dominant across nearly all encoder positions, while temporal cyclical variables such as \texttt{day\_of\_week\_cos} and progression indicators such as \texttt{time\_idx} maintain sustained influence throughout the forecasting window. The recency-effect analysis further confirms that the model prioritizes recent epidemiological context over distant historical information, while still preserving meaningful long-range temporal dependencies within the encoder sequence.

To better understand policy-sensitive behavior, we further analyzed SHAP responses for variables such as \texttt{stringency\_index}, \texttt{positive\_rate}, \texttt{new\_tests\_log}, and \texttt{new\_deaths\_log}. The resulting policy-impact visualizations demonstrated nonlinear relationships between intervention intensity and forecasted transmission risk. Such explanations are particularly useful for healthcare stakeholders and policymakers, as they help identify which intervention signals most strongly influence predicted outbreak trajectories and can therefore support evidence-driven public health planning.

\subsubsection{Public Health Interpretation}

Beyond predictive performance, the proposed SEIR-TFT framework provides interpretable insights that can support real-world public health decision-making. The SHAP-based attribution analysis revealed that historical infection trajectories, policy interventions, and SEIR latent states collectively play a major role in forecasting future outbreak dynamics.

For policymakers, the feature attribution results provide a quantitative understanding of how intervention-related variables such as \texttt{stringency\_index} influence transmission forecasts. The relatively high contribution of policy features (\(20.1\%\)) suggests that government restrictions and behavioral interventions substantially affect predicted epidemic trajectories. Such insights at each point in time can assist authorities in evaluating the effectiveness of containment strategies, identifying major outbreak drivers, and assessing the temporal impact of mobility restrictions or reopening policies.

For hospitals and healthcare systems, the strong temporal dependence learned by the TFT model enables improved anticipation of short-term case surges. Since the most influential signals were concentrated around recent infection windows (particularly near timestep \(t-5\)), the framework can support early operational planning for ICU utilization, staffing requirements, oxygen supply management, and emergency resource allocation during rapidly evolving outbreak periods.

For epidemiologists, the integration of SEIR latent states with deep temporal attention offers an interpretable bridge between mechanistic epidemic theory and data-driven forecasting. The substantial contribution of SEIR-derived variables (\(19.8\%\)) indicates that the learned representations remain epidemiologically meaningful rather than behaving as a purely black-box predictor. This alignment helps validate whether the learned temporal dynamics are consistent with expected transmission and recovery processes.

For regional health planners, demographic and static contextual variables such as \texttt{aged\_65\_older}, \texttt{diabetes\_prevalence}, and \texttt{cardiovasc\_death\_rate} provide additional insight into vulnerability patterns across countries. The interaction analysis suggests that population health conditions and intervention intensity jointly influence outbreak severity, enabling identification of high-risk regions that may require targeted surveillance, healthcare reinforcement, or prioritized vaccination efforts.

\subsection{Ablation Study}
To evaluate the contribution of the TFT (Temporal Fusion Transformer) component, a 
three-way comparison study was conducted. Outputs from pure SEIR mechanistic modelling, 
standalone TFT, and hybrid SEIR-TFT were evaluated against one another under identical training and evaluation settings. The SEIR dynamics 
were simulated using fixed epidemiological parameters consisting of transmission rate 
\(\beta=0.5\), incubation rate \(\sigma=\frac{1}{5}\), and recovery rate 
\(\gamma=\frac{1}{14}\), corresponding to an average incubation period of 5 days and 
infectious period of 14 days. The compartment transitions were then propagated forward 
through numerical integration of the SEIR differential equations.

Daily incident case counts in the SEIR baseline were approximated using the 
exposed-to-infectious transition term \(\sigma E_t\), which represents the number of newly 
infectious individuals generated at each forecasting step. To maintain consistency with the 
TFT forecasting target, SEIR-generated incidence estimates were transformed into the 
logarithmic space using:
\begin{equation}
    \hat{y}^{\mathrm{SEIR}}_t=\log(1+\sigma E_t)
\end{equation}
Forecasting was done on eight distinct locations selected according to geographic and 
population differences from the validation dataset. The results were directly compared 
against TFT predictions of \texttt{new\_cases\_log} using MAE, RMSE, and 
correlation-based evaluation metrics. Table~\ref{tab:ablation_results} summarizes the 
ablation study results.

\begin{table}[h]
    \caption{Ablation study comparing the standalone SEIR baseline, standalone TFT, and 
    the proposed SEIR-TFT hybrid model.}
    \centering
    \begin{tabular}{lccc}
        \toprule
        \textbf{Model} & \textbf{MAE} & \textbf{RMSE} & \textbf{Correlation} \\
        \midrule
        SEIR Baseline        & 1.4359 & 2.1860 & 0.0726 \\
        TFT Only             & 0.0381 & 0.1696 & 0.9852 \\
        SEIR-TFT Hybrid      & 0.0310 & 0.1600 & 0.9855 \\
        \bottomrule
    \end{tabular}
    \label{tab:ablation_results}
\end{table}

The three-way ablation reveals distinct contributions from each architectural component. 
The standalone SEIR baseline yields the weakest performance across all metrics, confirming 
that fixed mechanistic transitions with predetermined epidemiological parameters are 
insufficient to capture the heterogeneous, intervention-driven dynamics present in 
real-world pandemic data. The TFT-only model achieves competitive predictive performance, 
closely approaching the SEIR-TFT hybrid across all evaluation metrics, which reflects the 
strong capacity of attention-based temporal modelling to learn complex nonlinear 
dependencies directly from observed incidence sequences.

However, the hybrid outperforms the standalone TFT across MAE, RMSE, and 
correlation, demonstrating that the mechanistic SEIR prior contributes a statistically 
meaningful, if marginal, improvement. This finding suggests that while the TFT component 
is the dominant driver of predictive accuracy, the epidemiological structure encoded by 
the SEIR dynamics provides a complementary inductive bias that the data-driven component 
alone cannot fully recover. In particular, the SEIR prior anchors the model's behaviour 
during low-data regimes and at the onset of epidemic waves, where purely learned temporal 
dependencies are most susceptible to distributional shift. The hybrid formulation therefore 
cannot be reduced to a purely neural solution without measurable loss in forecasting 
fidelity, justifying the architectural complexity introduced by the mechanistic coupling \citet{syam2025graph}. In an experiment, the model also showed strong generalization across all pandemic splits even if it was trained on only a single split corresponding to the rising third-wave period, highlighting the robustness and transferability of the learned epidemic representations.

\section{Conclusion}

This study presented a hybrid SEIR--Temporal Fusion Transformer (SEIR-TFT) framework
for interpretable multi-region pandemic forecasting. By incorporating latent SEIR compartment states as time-varying inputs to the TFT, the proposed architecture injects
domain-grounded epidemiological structure directly into the sequence modeling pipeline,
rather than treating mechanistic and neural components as separate post-hoc ensembles.

The experimental evaluation across four phase-aware temporal splits demonstrated that
the hybrid framework substantially outperforms a fixed-parameter SEIR baseline across
all evaluation metrics. The three-way ablation study further established that while the
standalone TFT is a strong predictor in isolation, the integration of SEIR-derived
priors provides a consistent and meaningful improvement, particularly in capturing
epidemic onset dynamics and distributional shifts driven by policy interventions.

The incorporation of SHAP-based explainability analysis provided transparent,
feature-level attribution at both global and local levels. Historical infection
trajectories, government stringency signals, and SEIR latent states collectively emerged
as the dominant forecasting drivers, validating that the learned representations remain
epidemiologically meaningful rather than behaving as an opaque black-box predictor. These
interpretability outputs directly support targeted, actionable public health decision-making by
enabling policymakers, hospital administrators, and regional health planners to understand
\textit{why} forecasts change across time, place and people, not merely \textit{what} they predict.

Taken together, this work unifies mechanistic
epidemiology, transformer-based temporal learning, and explainable AI within a single
coherent framework oriented toward practical, policy-sensitive deployment.

\section{Limitations \& Future Work}

Despite the encouraging results, several limitations of the current framework warrant
acknowledgment. The approximations in SEIR calculation introduce structural
assumptions about incubation and infectious periods that may not generalize uniformly
across regions with different testing capacities, reporting delays, or healthcare
infrastructure. 

The SEIR component operates with fixed epidemiological parameters
\((\beta, \sigma, \gamma)\) that do not adapt to evolving transmission conditions such
as the emergence of new variants, shifting vaccine efficacy, or behavioral changes over
time. In practice, these parameters vary continuously across regions and pandemic phases,
and their static specification may limit the model's responsiveness during rapid epidemic
transitions.

The framework was evaluated exclusively on COVID-19 data. We selected COVID-19 because it provides one of the most comprehensive and consistently reported global infectious disease datasets, with extensive temporal, demographic, testing, and policy-related information available across a large number of countries. Despite its broad geographic coverage it reflects a single pathogen
with a specific transmission profile. Generalizability to other infectious diseases with
substantially different compartmental dynamics, such as seasonal influenza, Ebola, or
mpox, remains to be demonstrated.

SHAP explanations, while informative, are post-hoc and model-agnostic. They
approximate feature contributions over a sampled explanation set and may not perfectly
reflect the internal attention dynamics of the TFT under all forecasting conditions.
Intrinsic interpretability mechanisms, such as directly analyzing the TFT's variable
selection weights and temporal attention heads, could offer more architecturally faithful
explanations.

Finally, the current framework does not explicitly model spatial dependencies or
cross-country transmission linkages. Countries are treated as independent forecasting
units, which may underestimate the influence of international travel, regional spillover
effects, and synchronized wave dynamics on local outbreak trajectories.

Future research will focus on addressing these limitations in several directions. Adaptive and time-varying epidemiological parameters could be incorporated into the SEIR component to better capture changing transmission dynamics arising from new variants, vaccination coverage, and behavioral shifts. Evaluating the framework on additional infectious diseases with distinct transmission mechanisms would assess its generalizability beyond COVID-19. Spatial and mobility-aware representations has to be added which would enable modeling of cross-region transmission and spillover effects. Incorporating regional language support would facilitate stronger integration with the local healthcare ecosystem (\citet{10100598}, \citet{10.1007/978-3-032-05855-3_7}).As future dashboard interfaces evolve toward conversational decision support, healthcare specific guardrail mechanisms will also be necessary to ensure safe, polite, and ethically constrained interaction with public health stakeholders \citep{arun2025integrated}. Finally, future work will investigate more intrinsic interpretability approaches, including direct analysis of TFT variable-selection networks and attention weights, to complement post-hoc SHAP explanations and provide deeper insights into model decision-making. Future work may also explore multi agent LLM based policy simulation environments for modeling interactions among public health authorities, hospitals, and regional planners \citep{wang2026mindgames}, although such methods would require careful adaptation before use in epidemiological decision support.

\bibliographystyle{unsrtnat}
\bibliography{references}  






\section*{About the Authors}
\vspace{0.5em}

\noindent
\begin{tabular}{@{}p{0.18\textwidth}@{\hspace{0.02\textwidth}}p{0.80\textwidth}@{}}
    \raisebox{-0.85\height}{%
        \includegraphics[width=0.16\textwidth]{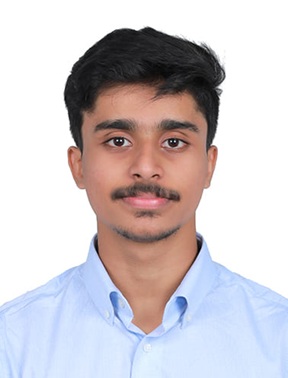}%
    } &
    \textbf{Ashish Anton Abraham} is an AI Engineer and part-time researcher affiliated with the Department of Computer Science and Engineering at the College of Engineering Trivandrum, Kerala, India. His research interests include deep learning, large language models, and artificial intelligence for edge and resource-constrained computing environments. His work focuses on the development of intelligent, interpretable, and efficient AI systems for real-world applications.

\end{tabular}

\vspace{1.0em}

\noindent
\begin{tabular}{@{}p{0.18\textwidth}@{\hspace{0.02\textwidth}}p{0.80\textwidth}@{}}
    \raisebox{-0.85\height}{%
        \includegraphics[width=0.16\textwidth]{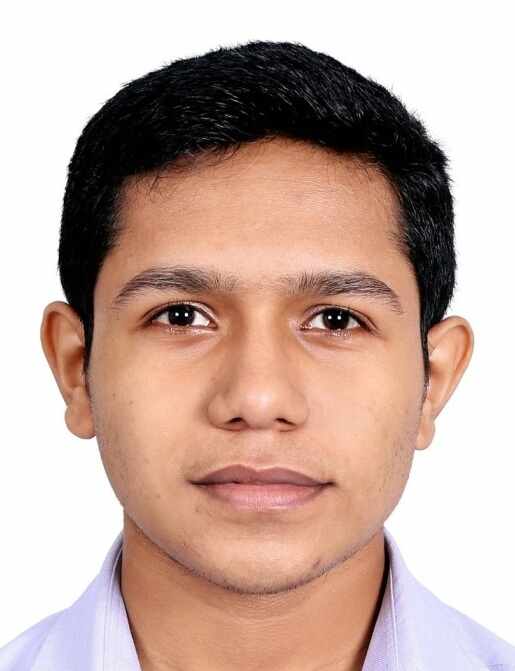}%
    } &
    \textbf{Gopi Krishnan R} is currently pursuing a Bachelor of Technology (B.Tech) degree in Computer Science and Engineering at the College of Engineering Trivandrum, Kerala, India. His research interests include epidemiological modeling, transformer architectures, explainable AI, and deep learning frameworks for healthcare.
\end{tabular}

\vspace{1.0em}

\noindent
\begin{tabular}{@{}p{0.18\textwidth}@{\hspace{0.02\textwidth}}p{0.80\textwidth}@{}}
    \raisebox{-0.85\height}{%
        \includegraphics[width=0.16\textwidth]{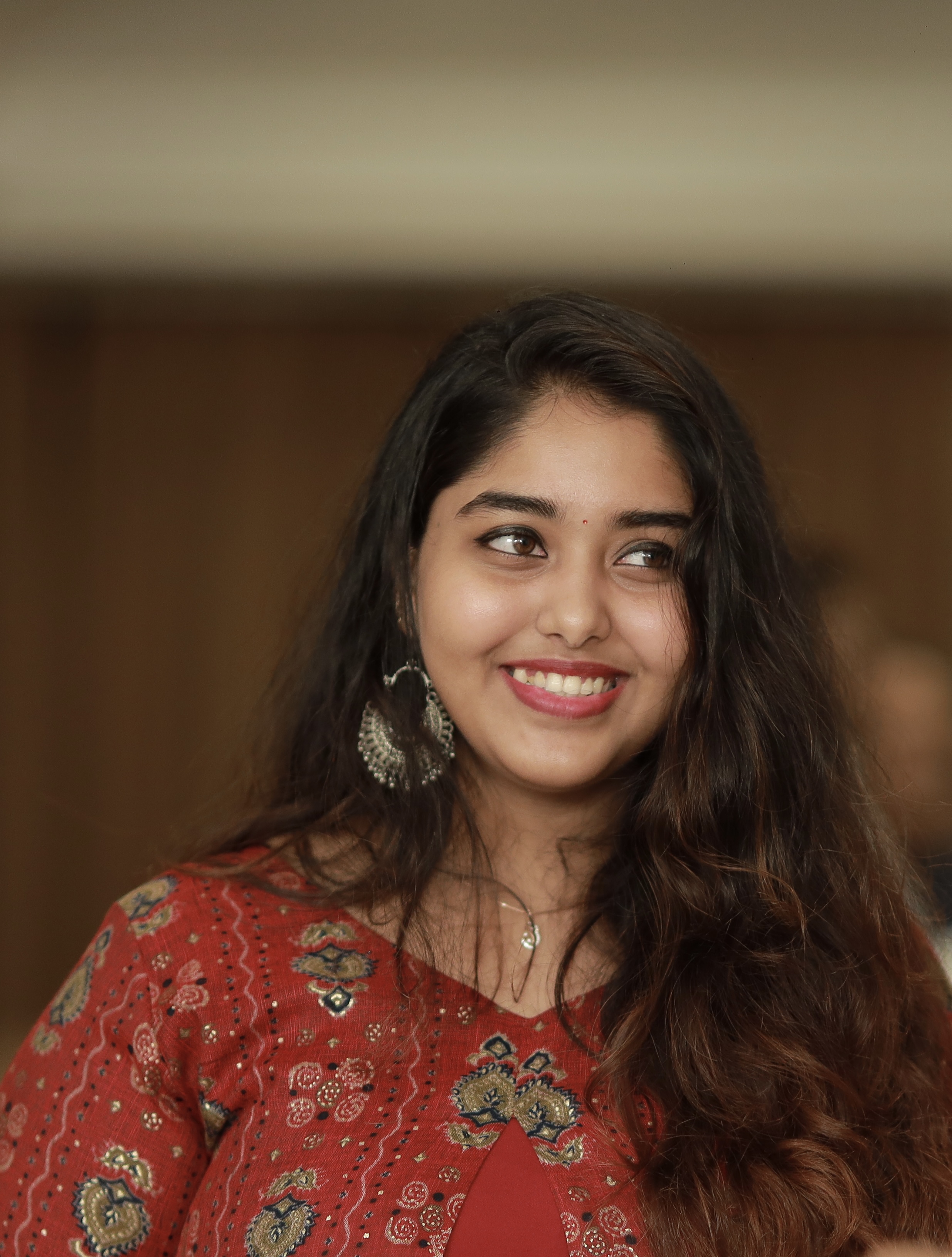}%
    } &
    \textbf{Aiswarya Anil Nair} is a postgraduate student at Indira Gandhi National Open University and a Machine Learning Engineer at O2V Private Limited, Chennai, India. Her work focuses on large language model systems, multi-agent architectures, and AI observability. Her research interests span natural language processing, generative AI, and responsible AI, with a particular emphasis on guardrails, safety mechanisms, and the real-world deployment of intelligent systems.
\end{tabular}

\vspace{1.0em}

\noindent
\begin{tabular}{@{}p{0.18\textwidth}@{\hspace{0.02\textwidth}}p{0.80\textwidth}@{}}
    \raisebox{-0.85\height}{%
        \includegraphics[width=0.16\textwidth]{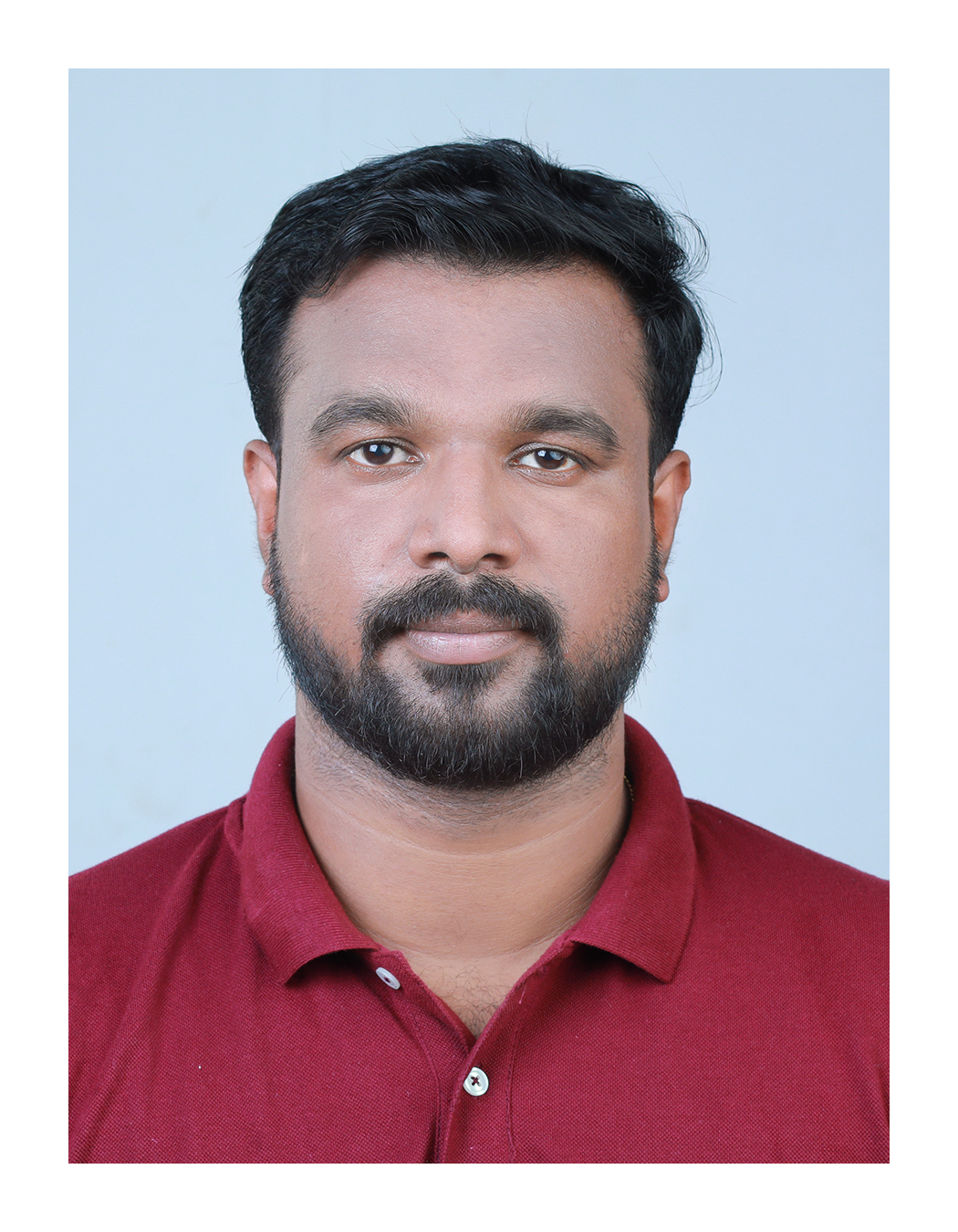}%
    } &
    \textbf{Dr. Ajo Babu George}  is a clinician who specializes in maxillofacial radiology and currently serves as the Founder and CEO of Medtech startup DiceMed. He works at the intersection of medical imaging and artificial intelligence. He received his graduation degree from Kerala University of Health Sciences and masters from Utkal University. His research areas include medical image segmentation and multimodal clinical reasoning.
\end{tabular}

\vspace{1.0em}

\noindent
\begin{tabular}{@{}p{0.18\textwidth}@{\hspace{0.02\textwidth}}p{0.80\textwidth}@{}}
    \raisebox{-0.85\height}{%
        \includegraphics[width=0.16\textwidth]{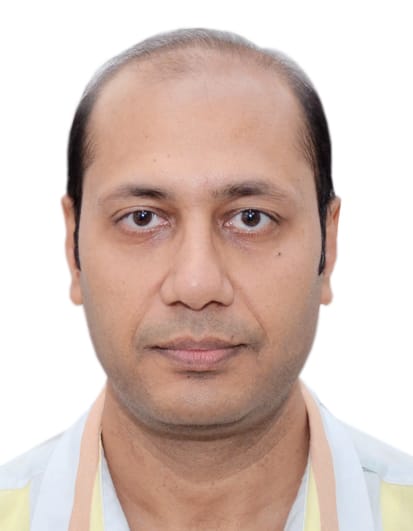}%
    } &
    \textbf{Dr. Kunal Agarwal} is a clinical track faculty at the Department of Oral Medicine and Radiology, SCB Dental College, Cuttack, Odisha. He is actively involved in research on the clinical validation of AI in dental imaging and explainable AI.
\end{tabular}
\end{document}